\documentclass[prd,superscriptaddress,twocolumn,10pt]{revtex4-1}
\usepackage{graphicx,color}
\usepackage{lipsum}
\usepackage{bm}
\usepackage{amssymb,amsmath,amsfonts, mathrsfs}
\usepackage{ulem} 

\newcommand{\be}{\begin{equation}}
\newcommand{\ee}{\end{equation}}
\newcommand{\ba}{\begin{array}}
\newcommand{\ea}{\end{array}}
\newcommand{\bqa}{\begin{eqnarray}}
\newcommand{\eqa}{\end{eqnarray}}

\begin{document}


\title{Reply to ``Comment on `Scrutinizing $\pi\pi$ scattering in light of recent lattice phase shifts'" }


\author{Xiu-Li Gao}
\affiliation{School of Physics, Southeast University, Nanjing 211189,
People's Republic of China}

\author{Zhi-Hui Guo}
\email[]{zhguo@seu.edu.cn}
\affiliation{School of Physics, Southeast University, Nanjing 211189,
People's Republic of China}

\author{Zhiguang Xiao}
\email[]{xiaozg@scu.edu.cn}
\affiliation{School of Physics, Si Chuan University, Chengdu~610065,
People's Republic of China}

\author{Zhi-Yong Zhou}
\email[]{Corresponding author: zhouzhy@seu.edu.cn}
\affiliation{School of Physics, Southeast University, Nanjing 211189,
People's Republic of China}
%
%


\maketitle
The commenting manuscript by Beveren and Rupp~(referred to as BR's comment~\cite{vanBeveren:2022zfx})
raises some doubts about our published
paper~\cite{Gao:2022dln}~(referred to as GGXZ paper in the following)
  on  the reliability and the novelty  of the results.

Before we reply to the comments, we need to highlight  some essential
points of GGXZ paper. In GGXZ paper, we have used the Peking
University~(PKU)  representation to
extract the information of the $S$-matrix poles from the lattice phase
shifts of the $\pi\pi$ scattering with $m_\pi=391,\ 236$ MeV by Hadron
Spectrum Collaboration
(HSC)~\cite{Briceno:2016mjc,Dudek:2012gj,PhysRevD.87.034505,Wilson:2015dqa}.
The starting point is that a reliable analysis should respect
the unitarity, crossing symmetry and analyticity of the scattering
amplitude.  In our analysis, the PKU representation respects the
analyticity and unitary, and the Balanchandran-Nuyts-Roskies~(BNR)
relation which relates all the three channels with $IJ=00,11,20$ in
the unphysical region between $s=0$ and $s=4m_\pi^2$ is also imposed to include the constraints from the crossing
symmetry. We stress that the BNR relation plays a much more relevant role in the present discussion when the $\sigma$ becomes a bound state below the two-pion threshold than the situation with a physical pion mass when the $\sigma$ is a broad resonance above the two-pion threshold.
By carefully comparing the different scenarios with and
without the BNR relation constraints, we concluded that the scenario with a virtual-state~(VS) pole
and a bound-state~(BS) pole in $IJ=00$ is better in describing
the lattice data and fulfill the BNR relations simultaneously than the one with only a $IJ=00$ BS pole. The scenario with only one BS pole can also reasonably describe the HSC phase-shift data. However, in this case the crossing
symmetry is badly violated. Therefore we would like to emphasize again that the crossing symmetry plays an important role in reaching our conclusion.

Their main criticism to our paper is that we  omit the contribution of
inelastic channels in our analysis and they argue that this is
important in affecting the existence of the virtual state. They use their own coupled channel model as the main tool to
support their argument. We would like to point out the differences between
their approaches and ours to clarify this issue.

Although the inclusion of
the dynamics above the inelastic channels may have some effect on the
determination of the $\sigma$ pole contents,
 our study reveals
that the crossing symmetry implemented via the BNR relation in the
unphysical energy region from $s=0$ to $s=4m_\pi^2$ is  more relevant
for investigation of the bound-state $\sigma$ below threshold than the
distant effects above the inelastic channels. The reason is simple:
the bound-state $\sigma$ is located inside the BNR integration ranges
and the inelastic channels are quite distant from such ranges.
Furthermore, we need to point out that a sophisticated model respecting the crossing
symmetry should at least have a left-hand cut in the partial wave
amplitude from the crossed $t$- and $u$-channel dynamics.
As far as we see, neither the effect of left-hand cut nor the
fulfillment of crossing symmetry for a large pion mass when $\sigma$
is turning into a bound state, are explicitly addressed in the BR's
commenting paper. In our opinion, the argument of $s$-$t$ channel
duality of the dual model in their comment
does not mean their model includes the crossing symmetry and there is no
analytical proof that their model satisfies the $s$-$t$ duality.
Including infinite
$s$-channel poles also does not mean that the model satisfies crossing
symmetry. A simple counterexample is that in the nonrelativistic
potential model, there could be infinite poles in the $s$ channel, but
we know that it does not have crossing symmetry.
  Also in our
opinion, being consistent with the phase shift data in the physical
region and $\sigma$ pole
position at physical pion mass is not enough to demonstrate the
consistency with the crossing symmetry and the $S$-matrix in the
unphysical region when pion mass is $391$MeV.
{{Furthermore, it is well-known that the negative phase shift of $IJ=20$
$\pi\pi$ channel is mainly contributed to by the left-hand cut from the
crossed channels, whereas a model
without the left-hand cut and crossing symmetry could hardly describe the
$IJ=20$ phase shift.}}
Thus, it is unclear to us for the large pion mass, when the $\sigma$ becomes a bound state below two-pion threshold whether their model could correctly
describe the $S$-matrix in the unphysical region between $s=0$ and $s=4m_\pi^2$, which is exactly where the BNR relations
are evaluated and the bound-state and virtual-state poles are located.
Regarding their statement that ``{\it However, for $m_\pi =
391$ MeV and with the fitted value of the overall coupling, we find a
bound state at 760 MeV, compatible with the lattice result of 758 MeV
in [2], but no nearby VS pole. This is to be contrasted to the
dispersive analysis in [1] for the same $m_\pi = 391$ MeV, which
extracted BS and VS poles at 1 MeV and 73 MeV below threshold,
respectively.}'', we have not seen any such documented analysis at
$m_\pi=391$MeV in their published papers.  From the BR's comment paper, we
also have not seen any numerical discussion about the left-hand cut and crossing symmetry in their model for the situation when $\sigma$ becomes a bound state.

\begin{figure}[h]%
\begin{center}%
\includegraphics[height=50mm]{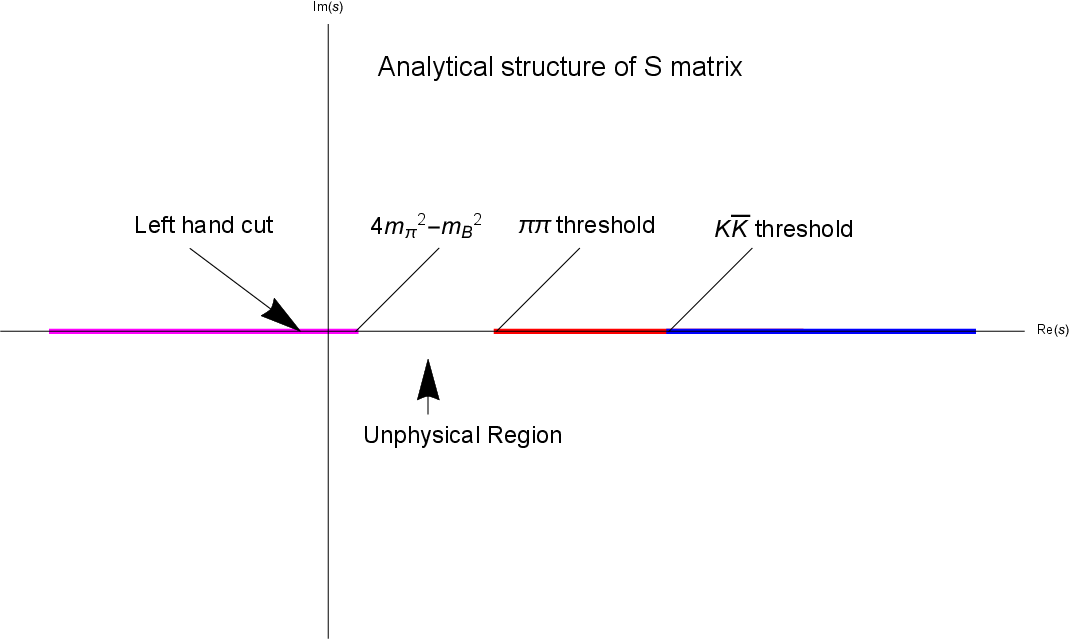}
\caption{\label{singularity} The cut structure of the $IJ=00$
$\pi\pi$ scattering amplitude when one of the $\sigma$ pole becomes a
bound state pole with $m_\pi=391$MeV and $m_K=549$MeV. There
appears another branch point at $4m_\pi^2-m_B^2$, which contribute another left-hand cut. }
\end{center}%
\end{figure}

To improve the GGXZ analysis for the purpose of including the $K\bar K$ contributions,
one needs to parametrize the contributions of right-hand cut integral and the higher poles including $f_0(980)$ pole, as we did in Ref.~\cite{Zhou:2006wm}. One needs to keep in mind that here we have $m_\pi=391$ MeV, and in this case,
according to the lattice result, $m_K=549$MeV and the $f_0(980)$ is found
to be at $1.166(45) - \frac i2\times 0.181(68)$GeV~\cite{Briceno:2017qmb}.
Now that our $S$-matrix is parametrized with respect to $s$, the
analytic structure of the $S$-matrix in the $s$ plane is shown in
Fig.\ref{singularity}.  We can see that the $K\bar K$ threshold
$\sim1.21$GeV$^2$  and $f_0$ pole at $s_{f_0}=1.35 - 0.21
i$GeV$^2$ are far from the $\pi\pi$ threshold $\sim 0.61$GeV$^2$.
The $K\bar K$ cut and the $f_0(980)$ is farther away from the
unphysical region than the left-hand cut, so they are not expected to affect much of the physics below the
$\pi\pi$ threshold.
As we have shown in the paper, since the BNR relation imposes
constraints in the whole unphysical region below the $\pi\pi$ threshold,
if only one $\sigma$ bound state is included then the $\chi^2$
contribution from BNR is large, whereas after the virtual state is included
the $\chi^2$ from BNR is significantly improved. This is the essential reason
why we include the virtual state. In order to quantitatively address
the concern of the BR's comment about the effects above the inelastic
$K\bar{K}$ channel, we have tried to explicitly include the $f_0(980)$
pole fixed by the lattice study~\cite{Briceno:2017qmb} in the PKU
parametrization. Within our expectation,
the $f_0(980)$ can not replace the virtual state to improve the $\chi^2$
significantly, and including both virtual state and the $f_0(980)$ does
not change the previous result much, since the $f_0(980)$ is far away from the
unphysical region between $s=0$ and $s=4m_\pi^2$ where the BNR relations are
evaluated.

Besides, in the original papers by the HSC group~\cite{Briceno:2016mjc,Dudek:2012gj,PhysRevD.87.034505,Wilson:2015dqa}, though the
 $s\bar
s$ operator is included  at $m_\pi=391$ MeV, the authors also stated that the
$K\bar K$ operators are not important
in determining the $\sigma$ pole
below the $\pi\pi$ threshold, which can be seen from their remark ``\emph {We
also include several $K\bar K$-like operators, of analogous
construction to the $\pi\pi$ operators, although they are not vital in
the determination of the spectrum below the $K\bar K$ threshold }".
In the updated paper
of HSC group~\cite{Briceno:2017qmb}, in order to include
the analysis of $f_0(980)$, they include the  whole $s\bar s$,  $K\bar
K$, $\eta\eta$ operators, etc.
The $\sigma$ pole position only moves from  $758(4)$ MeV to $745(5)$ MeV,
which also demonstrates that including the inelastic degrees of
freedom does not affect very much the physics below the $\pi\pi$ threshold.

Hereby, we clarify the author's confusion about our statement that ``\emph{The description of the $\sigma$ at
$m_\pi=391$~MeV as a pair of bound and virtual poles in our study is a
novel finding in the analysis of the lattice data, and to our knowledge
the role of the virtual state pole has not been explicitly reported in
other analyses of lattice data as in Refs. [41,42]}''.   Obviously, this statement
has two special premises: the first one is the special condition where  $m_\pi=391$MeV, and the second refers to
analyzing the specific lattice data set from the HSC.
To the best of our knowledge, we do not find such a conclusion in the previous publications when analyzing the HSC lattice data at $m_\pi=391$ MeV.
In such a special context, no one could assert that it is
a general result that there should be a virtual state there without a
careful study like we did.
The commenting authors also maintain that there is no virtual state in their
analysis, this also demonstrates that our result is different from
others, thus is novel in this special context.
Figure 1 in BR's
comment is  done by varying the coupling constant
$\lambda$.
However, Fig. 8 in GGXZ paper is given in a totally different
context, i.e. with varying pion mass. Though the figure in GGXZ looks similar
to the one in theirs, one can not
expect that it can be deduced from the behavior of varying coupling
constant, since the mass describes the intrinsic property of the
propagation of pion and has no direct relation with the coupling
constant. In our paper, we have
also explicitly mentioned that some other theoretical approaches also predicted
the $\sigma$ pole trajectory with respect to the change of pion masses: `` \emph {Such a pole behavior of the $\sigma$ resonance
has also been noticed in the calculations by unitarizing $\chi$PT
amplitude with a varying quark mass $m_q$ [9,44,45]}''

Finally, we take this opportunity to add a supplementary
discussion of the left-hand cut in our GGXZ paper. After the
publication, we realized that the left-hand cut discussed in our paper
is still not complete. In fact, after the $\sigma$ resonance becomes the
bound state at $s=m_B^2$ below the $\pi\pi$ threshold,
one more left-hand cut appears with a branching point at $4m_\pi^2-m_B^2$ which should also be
included in the BNR-relation, as shown in Fig.\ref{singularity}.
The omitted left-hand cut contributed by the bound-state $\sigma$ of
the crossed channel could be sizable since this cut overlaps with the energy region where BNR relation is evaluated.
Combining with the BNR relation, this extra left-hand cut effect could
be important in determining the fate of the virtual state revealed in our paper. No matter what direction the virtual-state pole finally goes to after the inclusion of the extra left-hand cut, it is mainly influenced by the crossing-symmetry requirement via the BNR relations, since the explicit inclusion of higher excited state $f_0(980)$ is verified to play little role in the BNR relation. We will
leave the detailed discussion about the effect of this extra left-hand cut to
a future work.

To conclude, the statement of the novelty of our result is clear: our conclusion about the $\sigma$ pole contents is given for a specific value of pion mass at
$m_\pi=391$MeV by simultaneously including the constraints from the lattice phase shifts and the crossing symmetry imposed by the BNR relations in the energy region between $0$ and $2m_\pi$.
Our study reveals that the left-hand cut and crossing symmetry play more important roles in determining the poles in unphysical region, instead of the distant effects lying above the inelastic channels, although the latter can affect the phase shifts to some extent in the elastic energy region. We have not
seen a numerical discussion of the crossing symmetry in the
commenting paper at large pion masses when $\sigma$ becomes a bound state below $\pi\pi$ threshold.
The insubstantial argument of
$s$-$t$ duality in
the commenting paper is not enough to prove the crossing
symmetry in their model.
As shown in Fig.\ref{singularity}, the left-hand cut is closer
to the unphysical region where the BNR relations take effect and
the $\sigma$ pole locates than the $K\bar K$
threshold and the $f_0(980)$ pole. To obtain reliable results
concerning the $\sigma$ state below the $\pi\pi$ threshold for large pion masses, the left-hand cuts and constraints of crossing symmetry below $\pi\pi$ threshold are important theoretical ingredients, which the BR's comment paper completely lacks. A numerical discussion of left-hand cut contribution and crossing symmetry
in their model is necessary if
they want to compare with our result of the dispersive approach,

\bibliographystyle{apsrev4-1}
\bibliography{ref}

\end{document}